\documentclass[a4paper,11pt]{article}

\usepackage{pos}

\usepackage{amsmath}
\usepackage{amssymb}
\usepackage{bm}
\usepackage{comment}
\usepackage{graphicx}
\usepackage{slashed}

\usepackage{xspace}
\usepackage{bbm}
\usepackage[binary-units=true]{siunitx}
\newcommand{\repoURL}{https://github.com/evanberkowitz/latex-base}

\newcommand{\Secref}[1]{Section~\ref{sec:#1}}

\newcommand{\figref}[1]{Fig.~\ref{fig:#1}\xspace}
\newcommand{\Figref}[1]{Figure~\ref{fig:#1}\xspace}
\renewcommand{\eqref}[1]{(\ref{eq:#1})\xspace}

\let\builtinLaTeX\LaTeX
\def\LaTeX{\builtinLaTeX\xspace}

\author*[a]{Finn Temmen}
\author[a]{Evan Berkowitz}
\author[c]{Anthony Kennedy}
\author[a,b]{Thomas Luu}
\author[b]{Johann Ostmeyer}
\author[c]{Xinhao Yu}

\affiliation[a]{Institute for Advanced Simulation (IAS-4),\\
Forschungszentrum Jülich, Germany}

\affiliation[b]{Helmholtz-Institut für Strahlen- und Kernphysik and Bethe Center for Theoretical Physics,\\
Rheinische Friedrich-Wilhelms-Universität Bonn, Germany}
	
\affiliation[c]{School of Physics and Astronomy, \\The University of Edinburgh, Scotland, UK}

\emailAdd{f.temmen@fz-juelich.de}
\emailAdd{e.berkowitz@fz-juelich.de}
\emailAdd{Tony.Kennedy@ed.ac.uk}
\emailAdd{t.luu@fz-juelich.de}
\emailAdd{ostmeyer@hiskp.uni-bonn.de}
\emailAdd{X.Yu-34@sms.ed.ac.uk}

\abstract{Despite its many advantages, the sensible application of the Hybrid Monte Carlo (HMC) method is often hindered by the presence of large - or even infinite - potential barriers.
These potential barriers partition the configuration space into distinct sectors, which leads to ergodicity violations and biased measurements of observables.
In this work, we address this problem by augmenting the HMC method with a multiplicative Metropolis-Hastings update in a so-called "radial direction" of the fields, which enables jumps over the aforementioned potential barriers at comparably low computational cost.
The effectiveness of this approach is demonstrated for the Hubbard model, formulated in a non-compact space by means of a continuous Hubbard-Stratonovich transformation.
Our numerical results show that the radial updates successfully resolve the ergodicity violation, while simultaneously reducing autocorrelations.}

\FullConference{The 41st International Symposium on Lattice Field Theory (LATTICE2024)\\
	28 July - 3 August 2024\\
	Liverpool, UK\\}

\title{Overcoming Ergodicity Problems of the Hybrid Monte Carlo Method using Radial Updates}
\ShortTitle{Overcoming Ergodicity Problems of the HMC Using Radial Updates}
\begin{document}
	
	\maketitle
	
	\section{Introduction}\label{sec:intro}
The Hybrid Monte Carlo (HMC) method \cite{HMC} is one of the most successful tools in the simulation of lattice field theories.
However, despite its many advantages, its application often faces challenges due to manifolds of vanishing fermion determinant and the concomitant emergence of infinite potential barriers.
These potential barriers lead to diverging force terms in the molecular dynamics (MD) evolution and, when separating regions in configuration space, result in an ergodicity violation of the algorithm.
A prime example for such a system is the Hubbard model, formulated in a non-compact space by means of a continuous Hubbard-Stratonovich (HS) transformation.
In the adopted formulation, the occurring fermion determinant vanishes on manifolds of codimension 1 and causes a formal ergodicity problem in HMC simulations \cite{UlybyshefPathIntRep}.
Thus, a sensible application of HMC necessitates the development of strategies to circumvent the potential barriers, which, in the context of the Hubbard model, has been discussed in great detail in \cite{Ergodicity_Hubbard}. 
In this work, we propose another method, which interleaves the HMC simulation with so-called \emph{radial updates}.
Radial updates refer to multiplicative Metropolis-Hastings (MH) updates in a radial direction of the non-compact fields, that enables jumps over the aforementioned potential barriers at comparably low computational cost.

In the following sections, we commence by introducing the Hubbard model in \Secref{HMC_Hubbard} and discuss ergodicity violations of the HMC due to the emergence of potential barriers.  
Afterwards, in \Secref{Radial}, we define the radial updates, before applying them to the simulation of the Hubbard model in \Secref{Results}. 
Finally, in \Secref{Summary}, we conclude with a summary and provide an outlook on potential future avenues.

	\section{Hubbard model}
\label{sec:HMC_Hubbard}

The Hubbard model is commonly used to describe strongly-correlated electrons in a variety of condensed matter systems.
We use the formulation of the Hubbard model in the so-called particle/hole basis, where the Hubbard Hamiltonian on a spatial lattice with $N_x$ sites is given by
\begin{align}
	\label{eq:Hubbard_Hamiltonian}
	H 
	= H_K + H_U 
	= -\kappa \sum_{\langle x, y\rangle} \left(
		  a_x^\dagger a_y^{\phantom{\dagger}}
		- b_x^\dagger b_y^{\phantom{\dagger}}
	\right)
	+ \frac{U}{2} \sum_x \left(
		  a_{x}^\dagger a_x^{\phantom{\dagger}} 
		- b_x^\dagger b_x^{\phantom{\dagger}}
	\right)^2. 
\end{align}
The Hamiltonian contains a nearest neighbor hopping term with the hopping parameter $\kappa$ and an on-site interaction with interaction strength $U$.
The fermionic operator $a_x^\dagger,$ ($a_x^{\phantom{\dagger}}$) creates (annihilates) a spin-$\uparrow$ electron at the lattice site $x$.
On the other hand, the operator $b_x^\dagger,$ ($b_x^{\phantom{\dagger}}$) creates (annihilates) a spin-$\downarrow$ electron-hole at the lattice site $x$. 
The partition function and expectation value of an observable $\mathcal{O}$ are given by the thermal traces
\begin{align}
	\label{eq:thermal_traces}
	Z 
	= \mathrm{tr}\left(
		e^{-\beta (H_K + H_U)}
	\right) 
	\quad \text{and} \quad 
	\langle \mathcal{O} \rangle 
	= Z^{-1} \mathrm{tr}\left( 
		\mathcal{O}e^{-\beta (H_K + H_U)} 
	\right). 
\end{align}
The application of the HMC method requires transitioning from the thermal traces \eqref{thermal_traces} to a path integral formulation, which is achieved by first discretizing the inverse temperature $\beta$ into $N_t$ time slices with lattice spacing $\Delta_t = \beta/N_t$ and performing a second order Suzuki-Trotter decomposition. 
This introduces an error of $\mathcal{O}(\Delta_t^2)$ and thus necessitates taking the continuum limit $N_t \rightarrow \infty$ to recover the exact expression. 
Applying a continuous HS transformation then decouples the many-body interactions at the cost of introducing a non-compact auxiliary bosonic field $\phi$. 
Finally, inserting fermionic coherent states and integrating out the fermionic degrees of freedom leads to the Hubbard action
\begin{equation}
	\label{eq:Hubbard_action}
	S[\phi] 
	= \frac{1}{2U\Delta_t}\sum_{t,x} \phi_{tx}^2 
	- \log \left( \det M[\phi|\kappa]\det M[-\phi|-\kappa]\right), 
\end{equation}
with the fermion matrix
\begin{equation}
	\label{eq:fermion_matrix}
	M[\phi|\kappa]_{tx,t'y} = 
	\delta_{t,t'}\delta_{x,y} 
	- \left(e^{\kappa h}\right)_{xy} e^{i\phi_{tx}}\mathcal{B}_{t'}\delta_{t',t+1}. 
\end{equation}
Here, $h = \Delta_t \delta_{\langle z, z'\rangle}$ is the hopping matrix and $\mathcal{B}_{t}$ encodes the anti-periodic boundary conditions with $\mathcal{B}_{t}=+1$ for $0<t'<N_t$ and $\mathcal{B}_{0}=-1$.
For a more detailed derivation of the Hubbard action \eqref{Hubbard_action} and similar formulations we refer the reader to \cite{Ergodicity_Hubbard,MCSimOfTightBindingGraphene}.

It was shown in \cite{UlybyshefPathIntRep,Ergodicity_Hubbard} that the fermion matrix \eqref{fermion_matrix} vanishes on manifolds with codimension 1, i.e.\@ on manifolds with dimension $d-1$ in the $d$-dimensional configuration space.
This gives rise to infinite potential barriers that separate regions in configuration space, and if the evolution of the molecular dynamics equations attempts to pass them, the force term $F[\phi] = -\frac{\partial S}{\partial \phi}$ diverges.
Therefore, the evolution is always repelled when approaching the potential barrier and the algorithm can not cross over into the separated region, resulting in an ergodicity violation. 
Thus, to obtain correct results, strategies for circumventing the potential barriers have to be developed and recent approaches are e.g.\@ the complex reformulation \cite{Revisiting_HQMC_for_Hubbard,HMCExtendedHubbardModel,AlgorithmForTheSimulationOfManyElectronSystems}, coarse MD integration \cite{Ergodicity_Hubbard}, using an operator with the same continuum limit but worse symmetry properties \cite{Ergodicity_Hubbard}, or utilizing novel generative machine learning architectures \cite{FlowsForHubbard}.  
In this work, we adopt the approach of using a non-Hamiltonian MH update to enable jumps over the potential barriers.
Specifically, we employ so-called radial updates, which will be introduced in the next section.

	\section{Radial Updates}
\label{sec:Radial}

Radial updates were proposed in Ref.~\cite{original_radial_update} and optimised in Ref.~\cite{cont_rad_up}. Here, they refer to a multiplicative Metropolis-Hastings (MH) update of a non-compact bosonic field $\phi = (\phi_1,\dots, \phi_d)$ that generates a new proposal by scaling the radius in field space
\begin{align}
	\label{eq:radius}
	R = \sqrt{\sum_{i = 1}^{d}\phi_i^2}. 
\end{align}
The radial updates are a special case of the updates used in more general multiplicative MH algorithms, such as the Random dive MH algorithm \cite{DuttaRDMH} and the Transformation-based Markov chain Monte Carlo method \cite{DuttaTMCMC}.
However, instead of devising a new multiplicative MH algorithm, we propose to augment a standard HMC simulation with intermediate radial updates.
This combined algorithm addresses the ergodicity problems of the standalone HMC at low computational cost, while also maintaining its favorable properties.
The radial update procedure is defined as follows:
\begin{itemize}
	\item[1. ] Given the initial configuration $\phi = (\phi_1,\dots, \phi_d)$, an update variable $\gamma$ is sampled from a zero-centered normal distribution $\mathcal{N}(\gamma |\mu = 0, \sigma_R^2)$ with standard deviation $\sigma_R^{\vphantom{2}}$. 
	\item[2. ] A new configuration $\phi'$ is generated by multiplying the initial configuration $\phi$ with $e^\gamma$, i.e.
	\begin{align}
		\phi' =  (e^\gamma\phi_1,\dots,e^\gamma\phi_d). 
	\end{align}
	This amounts to rescaling the radius of the initial configuration to $R'=e^\gamma R$, giving rise to the term \emph{radial update}. 
	The scaling factor $e^\gamma$ follows a log-normal distribution with a median of one, ensuring that increases and decreases in the radius are proposed with equal probability. 
	\item[3. ] The new configuration $\phi'$ is used as a trial configuration in a Metropolis acceptance test with acceptance probability
	\begin{align}
		\label{eq:radial_acc_prob}
		\alpha_{R} = \min\left(1, e^{-\Delta S + d\gamma }\right), 
	\end{align}
	where $\Delta S = S[\phi']-S[\phi]$ is the difference in action.
		Additionally, due to the multiplicative nature of the radial updates, the acceptance probability comprises a factor of $e^{d\gamma}$, stemming from the Jacobian of the transformation.
		Notably, in this setup, the acceptance probability is independent of the normal distribution used to sample $\gamma$, and therefore independent of $\sigma_R$ (except indirectly).
\end{itemize}
The radial updates satisfy the detailed balance condition, and therefore the combined algorithm of radial updates and HMC does as well. 
Furthermore, in the combined algorithm, the radial updates resolve possible ergodicity problems of the HMC by enabling jumps over potential barriers. 
In practice, employing radial updates requires additional design choices in comparison to a standalone HMC simulation.
These include selecting the standard deviation $\sigma_R$ and determining the frequency with which each type of step is applied.
Throughout this work, we will specify the latter by defining the ratio of HMC steps to the number of radial update steps in a combined update step.

	\section{Results}
\label{sec:Results}

In this section, we apply the combined algorithm of HMC and radial updates, as introduced in \Secref{Radial}, to simulate the Hubbard model \eqref{Hubbard_action}.
Specifically, we demonstrate the algorithm's effectiveness in overcoming ergodicity issues caused by potential barriers, as discussed in \Secref{HMC_Hubbard}. 
To achieve this, we examine the 2-site model, investigating autocorrelation times and the optimization of the additional parameter $\sigma_R$.
We then analyze how the method scales with growing dimensionality by increasing the number of time slices $N_t$. 
The simulations were performed using the \emph{Nanosystem Simulation Library} \cite{NSL} and the data analysis was performed using the \emph{comp-avg} tool \cite{comp-avg}. 

\subsection{Restoring ergodicity in the 2-Site model}
\label{sec:2Site}
We begin by examining a 2-site model on a single time slice, which allows for a direct visualization of the configurations $\phi = (\phi_1, \phi_2)$ and the ergodicity problems posed by the infinite potential barriers. 
We simulate the model using $U = 18$, $\kappa = 1$ and $\beta = 1$, employing both the standalone HMC and HMC augmented with radial updates. 
In this and in all following simulations, the MD integration is performed using the Leapfrog integrator with the trajectory length set to
\begin{align}
	\label{eq:optT}
	T=\frac{\pi}{2}\sqrt{U\beta/N_t}, 
\end{align}
which eliminates autocorrelations originating from the harmonic part of the action \cite{FourierAcceleration}.
Furthermore, throughout this work, the number of molecular dynamics steps is always tuned to obtain a fine integrator with acceptance rate $>99\%$, leading to $N_{\mathrm{md}}=60$ in the present simulation.
We record $N_{\mathrm{config}}=2\times 10^5$ configurations, saving the configuration after each HMC step.
In the simulation with radial updates, we employ one radial update per HMC step with standard deviation $\sigma_R = 1.84$.

\begin{figure}
	\centering
	\includegraphics[width = 0.9\textwidth]{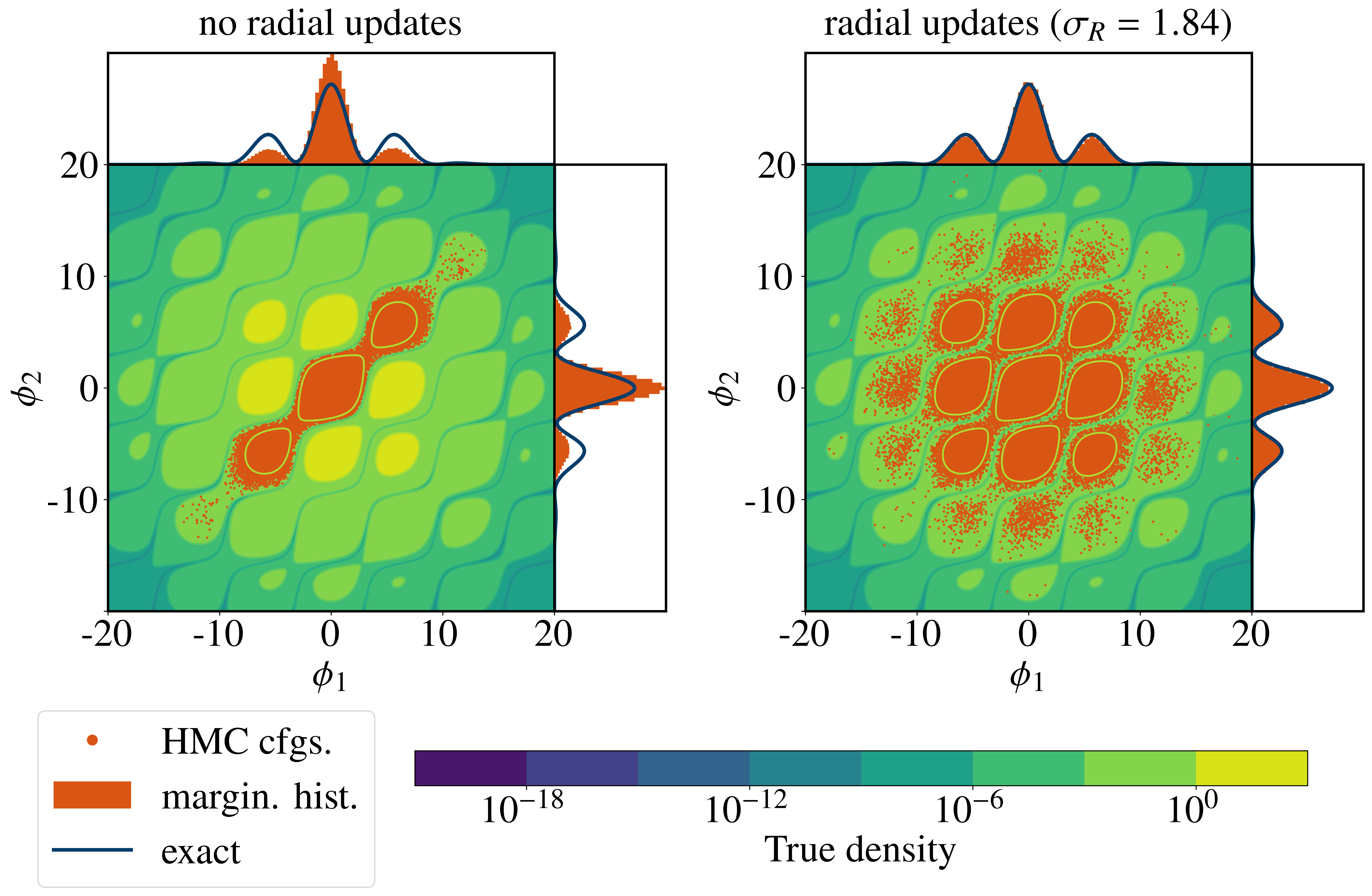}
	\caption{HMC configurations (dots) for the 2-site model on a single time slice ($N_t=1$) for $U = 18$, $\beta = 1$ and $\kappa = 1$.
	The configurations are compared to the exact unnormalized distribution, shown as a contour plot.
	The HMC simulations were conducted with $T=\frac{\pi}{2}\sqrt{U\beta/N_t}$ and $N_{\mathrm{MD}}=60$, achieving an acceptance rate $>99\%$.
	The left panel shows the simulation without radial updates, while the right panel includes radial updates.
	In the latter, a single radial update was performed before each HMC step with a proposal standard deviation of $\sigma_R = 1.84$.
	Each panel shows $10^5$ of the $N_{\mathrm{conf}}=2\times 10^5$ recorded configurations, with measurements taken after each HMC step.
	The plots at the margins compare the exact marginal distribution to the histograms obtained from the visualized trajectories, respectively.
	}
	\label{fig:Nt1_scatter}
\end{figure}

The first $10^5$ configurations in the Markov chain are visualized in \Figref{Nt1_scatter}.
Without radial updates, shown in the left panel, the HMC becomes trapped within the middle diagonal band in configuration space, which is separated from the adjacent bands by the aforementioned infinite potential barriers.
As a result, the simulation fails to sample regions of high probability in the adjacent bands, leading to a severe ergodicity problem. 
However, when radial updates are activated, as shown in the right panel of \Figref{Nt1_scatter}, the ergodicity violations are resolved entirely. 
This qualitative observation is further supported when histograms of the recorded configurations are compared to the exact marginal distributions, as depicted at the margins of the two plots in \Figref{Nt1_scatter}.
Without radial updates, the middle peak at $\phi_i=0$ is heavily oversampled in this example.
However, when radial updates are employed, the simulation closely aligns with the exact distribution.

\begin{figure}
	\centering
	\includegraphics[width = 0.9\textwidth]{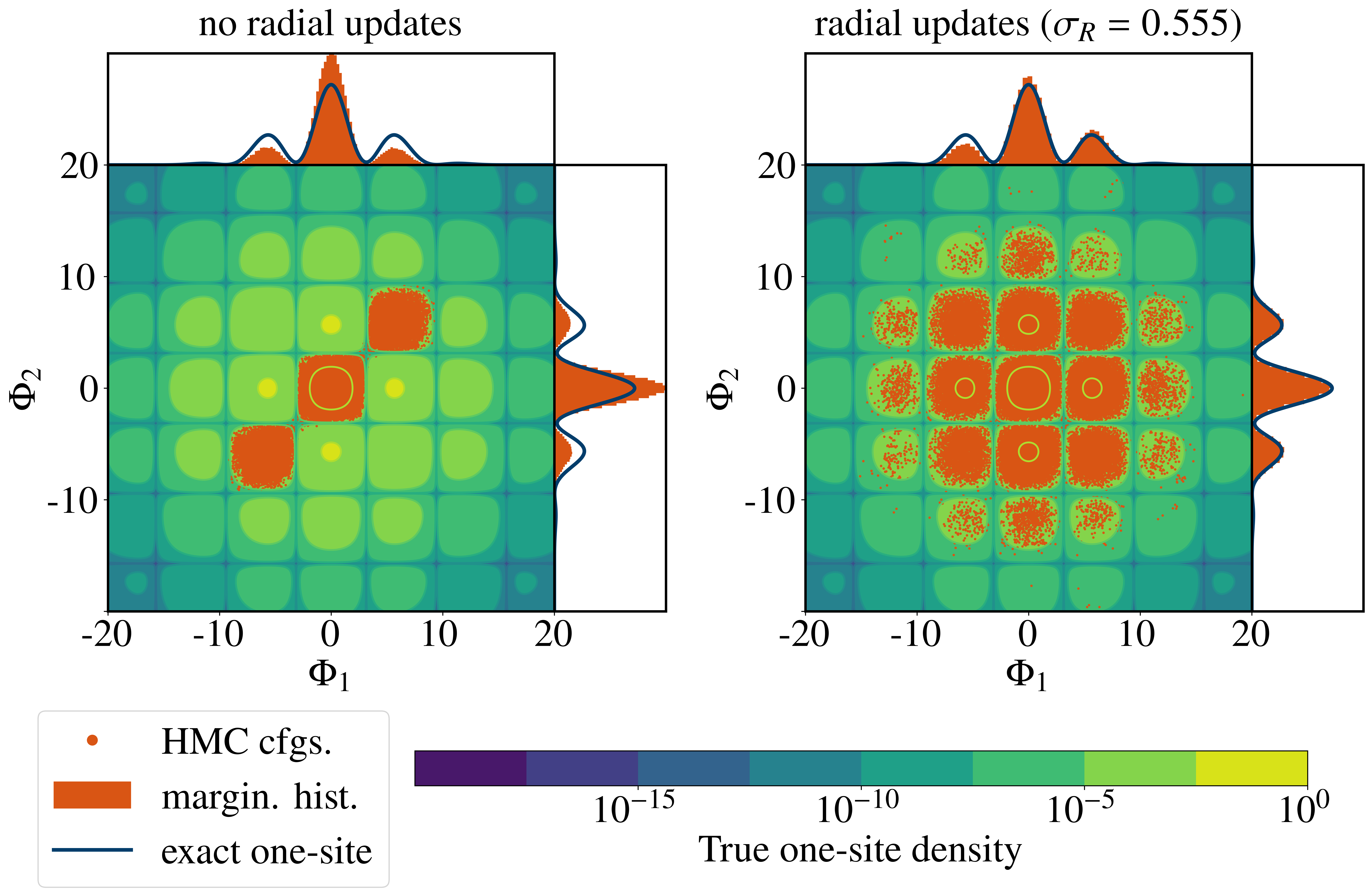}
	\caption{HMC configurations (dots) for the 2-site model with $N_t=8$ for $U = 18$, $\beta = 1$ and $\kappa = 1$.
	The configurations $\Phi=(\Phi_1, \Phi_2)$ with $\Phi_x = \sum_t \phi_{tx}$ are compared to the exact one-site distribution, recovered in the strong-coupling limit $U/\kappa \gg 1$ and shown as a contour plot.
	The HMC simulations were conducted with $T=\frac{\pi}{2}\sqrt{U\beta/N_t}$ and $N_{\mathrm{MD}}=60$, achieving an acceptance rate $>99\%$.
	The left panel shows the simulation without radial updates, while the right panel includes radial updates.
	In the latter, a single radial update was performed before each HMC step with a proposal standard deviation of $\sigma_R = 1.84$.
	Each panel shows $10^5$ of the $N_{\mathrm{conf}}=3\times 10^5$ recorded configurations, with measurements taken after each HMC step.
	The plots on the margins compare the marginal one-site distribution to the histograms obtained from the visualized trajectories, respectively.
	}
	\label{fig:Nt8_scatter}
\end{figure}

Next, we increase the dimensionality $d$ of the simple two-site model by increasing the number of time slices $N_t$, while keeping the parameters $U$, $\beta$ and $\kappa$ fixed. 
In this scenario, the potential barriers in configuration space can be visualized by considering $\Phi_x=\sum_t \phi_{tx}$, because in the strong-coupling limit $U/\kappa \gg 1$, the probability weights of a configuration $\Phi = (\Phi_1, \Phi_2)$ are well-approximated by the exact one-site distribution 
given in \cite{Ergodicity_Hubbard}. 
We first examine the exemplary case of $N_t = 8$ by visualizing the first $10^5$ of a total of $N_{\mathrm{conf}}=3\times 10^5$ recorded configurations, both without and with radial updates.
In the simulation with radial updates we employ one radial update with standard deviation $\sigma_R = 0.555$ per HMC step. 
The results are shown in \figref{Nt8_scatter}.
Similarly to the $N_t=1$ case, we observe that without radial updates, the trajectory remains trapped within the middle diagonal band in the two-dimensional $\Phi$-plane, whereas, with radial updates, the algorithm efficiently explores the entire configuration space, thereby restoring ergodicity. 

\subsection{Autocorrelations, parameter tuning and scaling}
\label{sec:Two_site_autocorr}
\begin{figure}
	\centering
	\includegraphics[width = 0.95\textwidth]{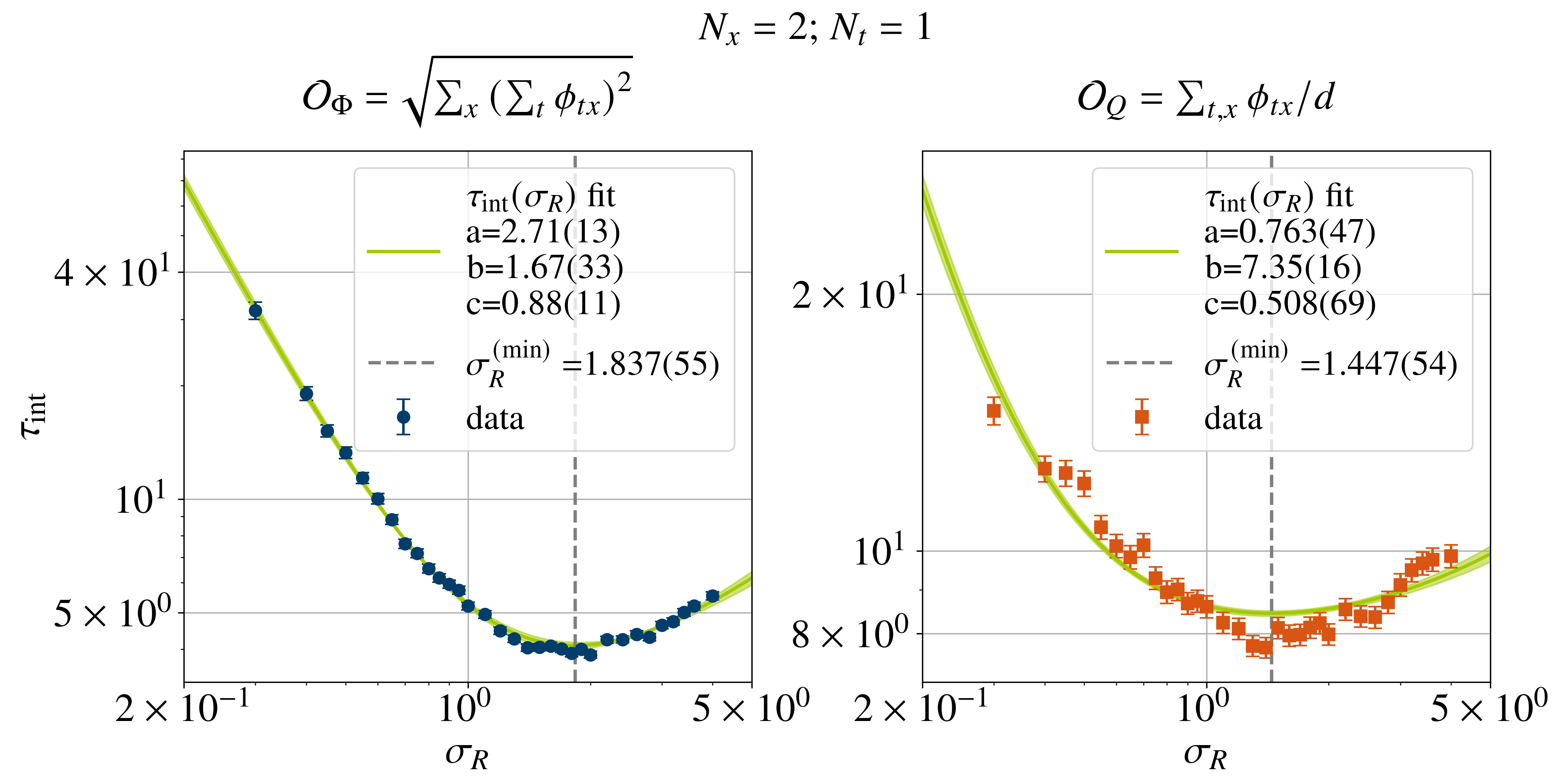}
	\caption{Integrated autocorrelation time $\tau_{\mathrm{int}}$ for the observables $\mathcal{O}_\Phi$ (left) and $\mathcal{O}_Q$ (right), defined in \eqref{observables}, as a function of the proposal standard deviation $\sigma_R$.
	The underlying model is the 2-site model on a single time slice ($N_t=1$) for $U = 18$, $\beta = 1$ and $\kappa = 1$.
	The HMC simulations were conducted with $T=\frac{\pi}{2}\sqrt{U\beta/N_t}$ and $N_{\mathrm{MD}}=60$, achieving an acceptance rate $>99\%$.
	Each HMC step was preceded by a single radial update and a total of $N_{\mathrm{conf}}=2\times 10^5$ configurations were recorded, with measurements taken after each HMC step.
	}
	\label{fig:R2S_Nt1_tauint}
\end{figure}

\begin{figure}
	\centering
	\includegraphics[width = 0.49\textwidth]{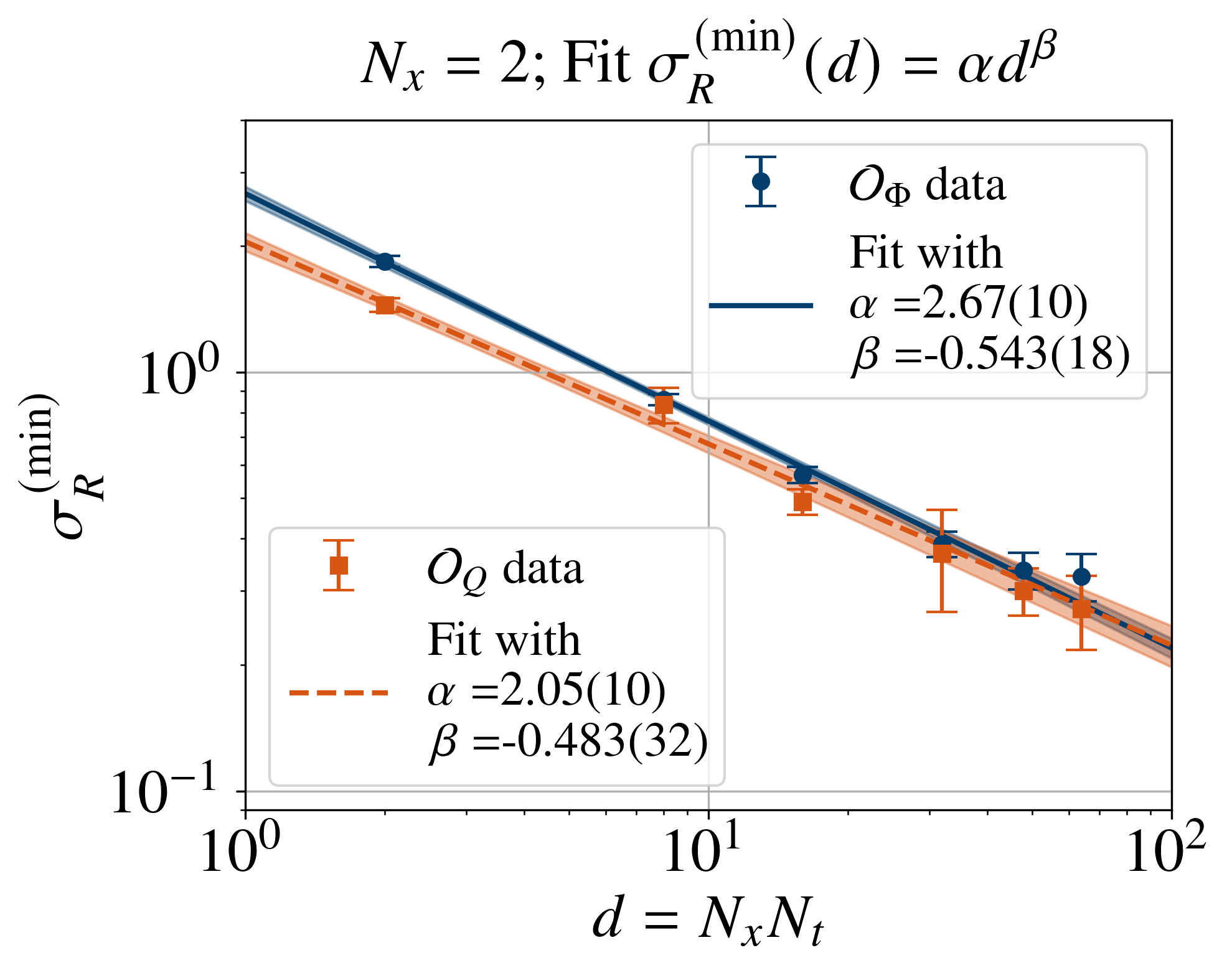}	
	\includegraphics[width = 0.49\textwidth]{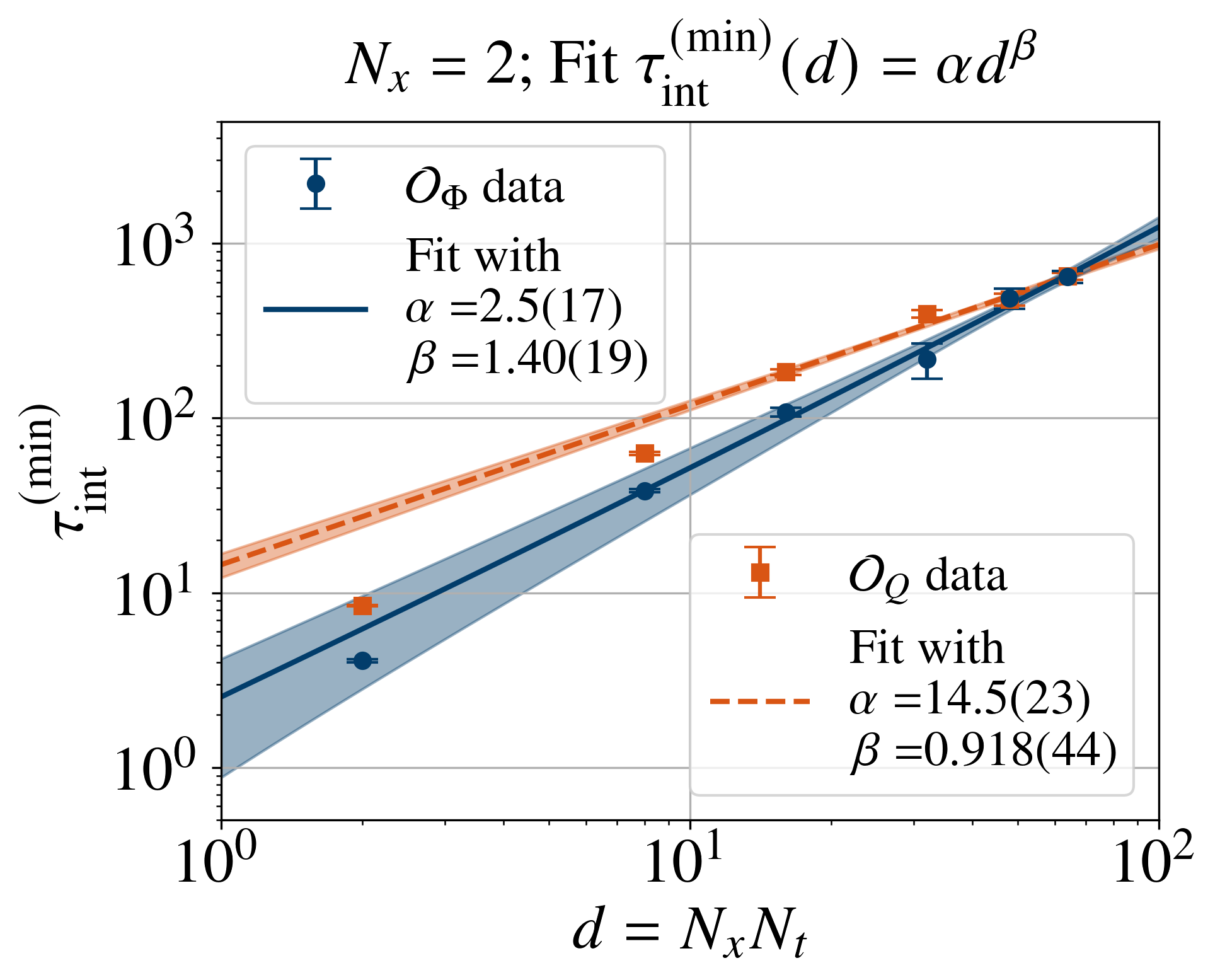}
	\caption{The left panel shows the position of the minimum $\sigma_R^{(\mathrm{min})}$ of the fit ansatz \eqref{tauint_fit} as a function of dimensionality $d=N_xN_t$, along with bootstrapped fits using the model $\sigma_R^{(\mathrm{min})}=\alpha d^\beta$.
	The right panel illustrates the integrated autocorrelation time at this minimum, denoted by $\tau_{\mathrm{int}}^{(\mathrm{min})}$, also as a function of dimensionality $d$, with corresponding bootstrapped fits using the model $\tau_{\mathrm{int}}^{(\mathrm{min})}=\alpha d^\beta$.
	Both panels display results for the 2-site model with varying $N_t$ for $U = 18$, $\beta = 1$ and $\kappa = 1$.
	}
	\label{fig:R2S_scaling}
\end{figure}

In the next phase of our analysis, we examine the influence of the radial updates on autocorrelations and utilize the integrated autocorrelation time to optimize the proposal standard deviation of the radial updates, while also exploring their scaling properties. 
We begin by computing the integrated autocorrelation time $\tau_{\mathrm{int}}$ for the 2-site model with $N_t=1$ as a function of the standard deviation $\sigma_R$.
For a comprehensive introduction to the estimation of errors and autocorrelation functions, we refer the reader to \cite{WolffAutocorr}.
The observables that we consider are the net charge of the system $\mathcal{O}_Q$ and the heuristically motivated $\Phi$-plane radius $\mathcal{O}_\Phi$, defined by 
\begin{align}
	\label{eq:observables}
	\mathcal{O}_Q = \frac{1}{d}\sum_{t,x}\phi_{tx}
	\quad\quad  \text{and} \quad\quad 
	\mathcal{O}_\Phi = \sqrt{\sum_x\left(\sum_t \phi_{tx}\right)^2}. 
\end{align}
The results are depicted in \figref{R2S_Nt1_tauint}, where we observe that $\tau_{\mathrm{int}}$ initially decreases as $\sigma_R$ increases, reaches a minimum, and then starts to rise.
To describe the dependence of the integrated autocorrelation time on the proposal standard deviation $\sigma_R$, we adopt the fitting ansatz
\begin{align}
	\label{eq:tauint_fit}
	\tau_{\mathrm{int}}(\sigma_R^{\vphantom{-2}}) = a\sigma_R^{-2} + b + c\sigma_R^{\vphantom{-2}}. 
\end{align}
In this expression, the first term accounts for the expected random walk behavior at small $\sigma_R$, resulting in a diffusive regime with $\tau_{\mathrm{int}}\propto \sigma_R^{-2}$. 
The third term quantifies the large $\sigma_R$ regime where proposed steps are large and the autocorrelation time increases with decreasing acceptance rate, such that $\tau_{\mathrm{int}}\propto \sigma_R^{\vphantom{-2}}$. 
The extrapolation is displayed in \figref{R2S_Nt1_tauint} and throughout this work, fit results are obtained by fitting the respective ansatz to $N_{\mathrm{boot}}=10^3$ bootstrap samples of the measured data.
The results clearly indicate that the chosen ansatz effectively captures the behavior of the data and, additionally, it enables the sensible estimation of the position of the minimum $\sigma_R^{(\mathrm{min})}$. 
To examine the scaling behavior of the radial updates, we conduct simulations across several values of $N_t$ and repeat the previous analysis. 
The respective estimates for $\sigma_R^{(\mathrm{min})}$ and the values for the integrated autocorrelation time at the minimum, denoted by $\tau_{\mathrm{int}}^{(\mathrm{min})}$, are shown in \figref{R2S_scaling}.
Theoretical considerations suggest that the position of the minimum should scale as $\sigma_R^{(\mathrm{min})}(d)\propto d^{-0.5} + \mathcal{O}(d^{-1})$ at leading order~\cite{cont_rad_up}, motivating the fit-model $\sigma_R^{(\mathrm{min})}(d) = \alpha d^\beta$. 
We find that the extrapolation of our numerical results closely matches the predicted  theoretical $d^{-0.5}$-scaling for both observables. 
To determine the leading order scaling of the minimal integrated autocorrelation time, we apply the same fit-model $\tau_{\mathrm{int}}^{(\mathrm{min})}(d) = \alpha d^\beta$, resulting in an almost linear leading order scaling of the minimal integrated autocorrelation time with the dimension $d$.

	\section{Summary and Outlook}
\label{sec:Summary}
In this work, we demonstrated that the augmentation of HMC with radial updates successfully resolves ergodicity violations of a standalone HMC in the Hubbard model.
Furthermore, the radial updates reduce autocorrelation times. We also discussed the scaling and tuning of additional parameters.
Due to the global multiplicative nature of the radial updates, their utilization is computationally inexpensive compared to the HMC itself.

In future research, we aim to investigate the scaling properties of radial updates further by extending our analysis to simulations of larger, more realistic systems, such as the simulation of Perylene \cite{RodekampPerylene}.
Additionally, performing a realistic simulation also involves utilizing a coarser MD integrator, which leads to energy violations in the integration that also enable tunnelling through potential barriers.
The employment of radial updates also has great implications for the geometrical convergence of HMC on non-compact manifolds, which will be subject to future work.

	\begin{acknowledgments}
	This work was funded in part by the Deutsche
Forschungsgemeinschaft (DFG, German Research Foundation) as part of the CRC 1639 NuMeriQS–project no.~511713970.
	We gratefully acknowledge the computing time granted by the JARA Vergabegremium and provided on the JARA Partition part of the supercomputer JURECA at Forschungszentrum Jülich~\cite{jureca-2021}.
	\end{acknowledgments}

	
\end{document}